\documentstyle[11pt]{article}

\def\g{\gamma}
\def\d{\delta}

\def\f{\phi}
\def\vf{\varphi}
\def\la{\lambda}

\def\p{\pi}

\def\cent{\centerline}
\def\pa{\parindent}

\def\vs{\vskip}
\def\ni{\noindent}
\def\pa{\parindent}
\def\ej{\vfill\eject}

\def\ifmath#1{\relax\ifmmode #1\else $#1$\fi}%
\def\Pp{\ifmath{{\mathrm{p}}}}
\def\PK{\ifmath{{\mathrm{K}}}}
\def\ra{\ifmath{{\mathrm{a}}}}

\def\rb{\ifmath{{\mathrm{b}}}}
\def\rc{\ifmath{{\mathrm{c}}}}
\def\rd{\ifmath{{\mathrm{d}}}}
\def\re{\ifmath{{\mathrm{e}}}}
\def\rf{\ifmath{{\mathrm{f}}}}

\def\rt{\ifmath{{\mathrm{t}}}}

\def\ty{\ifmath{{\mathrm{ty}}}}
\def\tz{\ifmath{{\mathrm{tz}}}}

\def\LAB{\ifmath{{\mathrm{LAB}}}}

\newcommand{\beqa}{\begin{eqnarray}}
\newcommand{\eeqa}{\end{eqnarray}  }
\newcommand{\beqan}{\begin{eqnarray*}}
\newcommand{\eeqan}{\end{eqnarray*}}
\newcommand{\beq}{\begin{equation}}
\newcommand{\eeq}{\end{equation}  }

\begin{document}


{\Large Self-affine fractality in $\p^+$p and K$^+$p collisions at 250 GeV/$c$}
\vs 1cm
\cent{EHS/NA22 Collaboration}
\vs 1cm
{\pa=0pt
N.M. Agababyan$^{8}$, 
M.R. Atayan$^{8}$, 
M. Charlet$^{4,\ra}$, 
J. Czy\.{z}ewski$^{4,\rb}$,
E.A. De Wolf$^{1,\rc}$, \\
K. Dziunikowska$^{2,\rd}$, 
A.M.F. Endler$^{5}$,
Z.Sh. Garutchava$^{6}$, 
H.R. Gulkanyan$^{8}$,\\
R.Sh. Hakobyan$^{8}$,
J.K. Karamyan$^{8}$, 
D. Kisielewska$^{2,\rd}$,
W. Kittel$^{4}$, 
L.S. Liu$^{7}$,\\
S.S. Mehrabyan$^{8}$,
Z.V. Metreveli$^{6}$, 
K. Olkiewicz$^{2,\rd}$, 
F.K. Rizatdinova$^{3}$, 
E.K. Shabalina$^{3}$,\\ 
L.N. Smirnova$^{3}$,
M.D. Tabidze$^{6}$, 
L.A. Tikhonova$^{3}$,
A.V. Tkabladze$^{6}$,
A.G. Tomaradze$^{6,\re}$,\\
F. Verbeure$^{1}$,
Y.F. Wu$^{4,\rf}$,
S.A. Zotkin$^{3}$

\begin{itemize}
\itemsep=-2mm
\labelsep=2mm
\itemindent=-2mm
\item[$^{1}$] Department of Physics, Universitaire Instelling Antwerpen,
B-2610 Wilrijk, Belgium
\item[$^{2}$] Institute of Physics and Nuclear Techniques of Academy of
Mining and Metallurgy and Institute of Nuclear Physics, PL-30055 Krakow,
Poland
\item[$^{3}$] Nuclear Physics Institute, Moscow State University, 
RU-119899 Moscow, Russia
\item[$^{4}$] High Energy Physics Institute Nijmegen (HEFIN), University of 
Nijmegen/NIKHEF, NL-6525 ED Nijmegen, The Netherlands
\item[$^{5}$] Centro Brasileiro de Pesquisas Fisicas, BR-22290 
Rio de Janeiro, Brazil
\item[$^{6}$] Institute for High Energy Physics of Tbilisi State 
University, GE-380086 Tbilisi, Georgia
\item[$^{7}$] Institute of Particle Physics, Hua-Zhong Normal 
University, Wuhan 430070, China
\item[$^{8}$] Institute of Physics, AM-375036 Yerevan, Armenia
\end{itemize}
 
\vs 1cm
{\bf Abstract:}
Taking into account the anisotropy of phase space in multiparticle 
production, a
self-affine analysis of factorial moments was carried out  
on the NA22 data for $\p^+\Pp$ and $\PK^+\Pp$
collisions at 250 GeV/$c$. Within the transverse
plane, the Hurst exponents measuring the anisotropy
are consistent with unit value (i.e. no anisotropy). They are, however, 
only half that value when the longitudinal
direction is compared to the transverse ones. Fractality, indeed,
turns out to be self-affine rather than self-similar in 
multiparticle production. In three-dimensional
phase space, power-law scaling is observed to be better realized in 
self-affine than in self-similar analysis. 


\vfill
\hrule width3truecm
\vs 2mm
$^\ra$ EC guest scientist, now at DESY, Hamburg\\
$^\rb$ KUN Fellow from the Jagellonian University and fellow of the Polish
Science Foundation (FNP) scholarship for the year 1996, Krakow\\
$^\rc$ Onderzoeksleider NFWO, Belgium\\
$^\rd$ Supported by the Polish State Committee for Scientific Research\\
$^\re$ Now at UIA, Wilrijk, Belgium\\
$^\rf$ KNAW visitor from Hua-Zhong Normal University, Wuhan, China
\par}

\ej
\section{Introduction}

The suggestion that normalized factorial moments $F_q(\d y)$ of 
particle-multiplicity distributions in ever smaller phase-space intervals 
$\d y$, may show power-law behavior [1] 
\beq
F_q(\d y) \propto (\d y)^{-\f_q}\ \ , \quad \quad (\d y\to 0)
\eeq
has spurred a vigorous experimental search for linear dependence of $\ln F_q$
on $-\ln \d y$ [2]. Analogous to a similar behavior at the onset of
turbulence, such a dependence is referred to as ``intermittency''.
Power-law scaling is typical for fractals [3], i.e.,
for self-similar objects of non-integer dimension. The powers $\f_q$ 
are related to the anomalous dimensions $d_q=\f_q/(q-1)$ measuring the 
fractality of a system [4]. In general, however, only approximate scaling 
has been observed in the experimental data.

When comparing log-log plots for one phase-space dimension, one notices
that the ln$F_q$ saturate at small $\d y$. This can be explained as a
projection effect of a three-dimensional phenomenon [5].  In 
three-dimensional analysis, however, the power law also does not hold 
exactly in all data. 
In NA22, for example, the 3D data are seen to even bend upward [6].

A deviation from exact scaling can be expected from an anisotropy of
occupied phase-space.  To account
for such an anisotropy, Wu and Liu have suggested [7] that the scaling 
property should be different in 
longitudinal and transverse directions
and the local multiplicity fluctuations are {\em self-affine}
rather than {\em self-similar}.
If that is the case, the anomalous scaling of factorial moments can be
observed to be retained exactly, only under a self-affine analysis,
where the shrinking ratio is allowed to be different in 
different directions. 
In a self-similar analysis, in contrast, all  
directions are forced to have an identical shrinking ratio.
It should be remembered, however, that the scaling law is expected
to be distorted in full phase space due to correlations imposed by
momentum conservation.

The experimental data sample is described in Sect.~2. The method of self-affine
analysis is briefly summarized in Sect.~3. The results of the self-similar 
analysis are shown in Sect.~4. After reducing
the influence of momentum conservation in the full region, so-called Hurst 
exponents are obtained by fitting one-dimensional factorial moments as
shown in Sect.~5.
In Sect.~6, the results from the self-affine analysis compared to those
from the self-similar analysis are given. Conclusions are summarized in Sect.~7. 

\section{The data sample}

In the CERN experiment NA22, the European Hybrid Spectrometer (EHS) was 
equipped with the Rapid Cycling Bubble Chamber (RCBC) as an active target
and exposed to a 250GeV/$c$ tagged, positive, meson enriched beam. In data
taking, a minimum bias interaction trigger was used. The details of 
spectrometer and trigger can be found in [8,9].

Charged particle tracks are reconstructed from hits in the wire- and
drift-chambers of the two-lever-arm magnetic spectrometer and from measurements
in the bubble chamber. The momentum resolution varies from 1-2\% for tracks
reconstructed in RCBC, to 1-2.5\% for tracks reconstructed in the first
lever arm and is 1.5\% for tracks reconstructed in the full spectrometer.

Events are accepted for the analysis when the measured and reconstructed charge
multiplicity are the same, charge balance is satisfied, no electron is detected
among the secondary tracks and the number of badly reconstructed (and therefore
rejected) tracks is 0. The loss of events during measurement and reconstruction
is corrected for by applying a multiplicity-dependent event weight 
normalized to the  
topological cross sections given in [9]. Elastic events
are excluded. Furthermore, an event is called single-diffractive and excluded
from the sample if the total charge multiplicity is smaller than 8 and at least
one of the positive tracks has a Feynman variable 
$|x_{\ifmath{{\mathrm{F}}}}|>0.88$. After these cuts, the inelastic,
non-single-diffractive sample consists of 59 200 $\pi^+\Pp$ and
$\PK^+\Pp$ events.

For laboratory-momenta $p_{\LAB}<0.7$GeV/$c$, the range
in the bubble chamber and/or the change of track curvature is used for proton
identification. In addition, a visual ionization scan is used for
$p_{\LAB}<1.2$GeV/$c$ on the full $\PK^+\Pp$ and on 62\% of the
$\pi^+\Pp$ sample. Positive particles with $p_{\LAB}>150$GeV/$c$ are given 
the identity of the
beam particle. Other particles with momenta $p_{\LAB}>1.2$ GeV/$c$ are not
identified in the present analysis and are treated as pions.

In spite of the electron rejection mentioned above, residual Dalitz decay
and $\g$ conversion near the vertex still contribute to the two-particle
correlations. Their influence on our results has been investigated in detail
in [10].

\section{The method}

In the language of self-affine analysis in three-dimensional phase space
(here denoted as $p_a, p_b, p_c$), only under the self-affine transformation
$\d p_a \to \d p_a/\la_a,\ \d p_b \to \d p_b/\la_b,\
\d p_c \to \d p_c/\la_c\ $ with non-identical shrinking ratios $\la_a$, 
$\la_b$ and $\la_c$, are the factorial moments expected to
have the well-defined scaling property

\begin{equation}
F_q(\d p_a, \d p_b, \d p_c)= \la_a^{\phi_q^{(a)}}\la_b^{\phi_q^{(b)}}
\la_c^{\phi_q^{(c)}} F_q(\la_a \d p_a, \la_b \d p_b, \la_c \d p_c).
\end{equation}

\noindent The shrinking ratios $\la_a,\la_b,\la_c$ are
characterized by the so-called roughness or Hurst exponents [11]
\begin{equation}
H_{ij}={\ln \la_i\over \ln \la_j},\ \ \ \ \ \ \  (i,j=a,b\quad \mbox{or} \quad
 a,c\quad
\mbox{or} \quad b,c\quad ),  \end{equation}

\noindent with
\begin{equation}
\la_i \le \la_j, \quad \quad 0\le H_{ij}\le 1, \end{equation}
describing the anisotropy of the system under study. 
For $H_{ij}=0$, $\la_i=1$, the scaling property does not exist in 
direction $i$, only in direction $j$.
For $H_{ij}=1$, the self-affine transformation reduces to a 
self-similar one,
meaning that the system is isotropic in these two directions. For $0 < H_{ij}
< 1$, non-trivial self-affine fractality exists in the $(i,j)$ plane, i.e, the
fluctuation is anisotropic in that plane.

The Hurst exponents can be deduced from the data by
fitting three one-dimensional second-order factorial-moment saturation 
curves [5]
\begin{equation}
F^{(i)}_2(M_i)=\alpha_i-\beta_iM_i^{-\gamma_i}, \qquad \qquad   (i=a,b,c)
\end{equation}
where $M_i=\Delta p_i/\delta p_i$ is the number of sub-divisions in 
direction $i$, $\Delta p_i$ and $\delta p_i$ are the initial and 
final interval size in direction $i$, respectively, and $\alpha_i$, 
$\beta_i$ and $\gamma_i$ are
three fit parameters. The Hurst exponents are determined from the parameter
$\gamma_i$ as 

\begin{equation}
H_{ij}={1+\gamma_j\over 1+\gamma_i},\qquad  \qquad (i,j=a,b\quad \mbox{or} 
\quad a,c\quad \mbox{or}\quad b,c).
\end{equation}
\noindent
With these Hurst exponents, a self-affine analysis
can be executed according to (2). If self-affine fluctuations of multiplicity 
do exist in multiparticle production, exact scaling should be observed
in three-dimensional phase space.

A scaling function similar to (2) for two variables has also been
suggested by J. Wosiek [12] as a requirement for hyperscaling
from a formal analogy with statistical physics.

\section{Self-similar analysis}

The results of a {\it self-similar} analysis in 1-, 2- and 3-dimensional 
phase space are presented in Fig.~1. 
The initial intervals for the three phase-space variables, rapidity $y$, 
azimuthal angle $\vf$ and transverse momentum $p_{\rt}$, are defined as:

$$-2<y<2$$
$$0<\vf\le 2\pi$$
$$0.001<p_\rt<10 \ {\mbox{GeV/}}c. $$


\noindent
To avoid trivial effects due to lack of translational invariance, all 
variables are transformed to the corresponding cumulative variables
by the Ochs method [13]. The experimental resolution in $y$, $\vf$ and
$p_\rt$
has been studied in detail in [10]. The limited available statistics,
rather than the experimental resolution, 
sets the limit on the smallest bin size to be used in the analysis.

In one-dimensional projection, the 
partitioning $M=1,2,\dots 40$ is used for 
all three variables. In two-dimensional projection, the partitioning in each 
direction is $M=1,2,\dots 20$, so that for the area it is $M_2=1^2,2^2,\dots 
20^2$. In the three-dimensional case, $M=1,2,\dots 15$ is used for each 
direction, so that $M_3=1^3,2^3,\dots 15^3$ for a three-dimensional box.


In the 1-D analysis (first column of Fig.~1), $F_2$ saturates at three 
different values when using
$y$, $p_\rt$ and $\vf$, respectively. 
In the case of $y$, $F_2$ increases rapidly with increasing $\ln M$ at 
small $M$ and reaches a saturation value which is the highest of the three. 
The trend is followed for 
the case of $p_\rt$, but at lower values of $F_2$. When using $\vf$, on the 
other hand, $F_2$ increases with increasing $\ln M$ only above an initial
decrease.

From the 2-D analysis (second column of Fig.~1) in the $(y,p_\rt)$
plane, we observe an onset for a saturation at medium $\ln M$, followed
by an upward bending at large $\ln M$. An upward bending is observed 
in the $(y,\vf)$ plane. In the $(p_\rt, \vf)$ plane, a decrease at low 
$\ln M$ is followed by an upward bending.  



  
In the 3-D analysis, ln$F_q$ is bending upward for all orders of $q$.


\section{Hurst exponents for higher-dimensional phase space}

Momentum conservation by itself causes a correlation and can, therefore,
distort the scaling behavior expected from the dynamics of particle
production [14]. The influence of momentum conservation on the factorial
moments is expected to be different in the various 
variables. The variable $p_\rt$ contains only 
the absolute value of momentum in the transverse plane without any 
information on the direction. The influence of momentum conservation,
therefore, 
is small for this case. For rapidity $y$, the influence of leading particles 
is reduced by the $y$-cut given above. Therefore, the influence of momentum 
conservation in $y$ is not significant. For the variable $\vf$,
however, all directions in the transverse plane 
are included for $M=1$, so that $F_2$ is dominated 
by transverse momentum compensation,
which explains the decrease of $F_2(\vf)$  with increasing $\ln M$ at low 
$\ln M$ as shown in Fig.~1.




After reducing the influence of momentum conservation by
excluding low values of $M$, it is easy to obtain the Hurst exponents from
the data by means of (5) and (6). The fit results obtained according to 
(5) are shown in Fig.~2 
for all phase-space variables considered. The parameter values are given in 
Table 1. Accordingly, the Hurst exponents deduced from (6) are:

$$H_{yp_\rt} = {1.021\over 2.14}=0.48 \pm  0.06 \ ;$$
$$H_{y\vf} = {1.0139 \over 2.14}=0.47 \pm 0.06 \ ;$$
$$H_{p_\rt\vf} ={1.014\over 1.0212}= 0.99 \pm 0.01 \ .$$
\noindent
From these Hurst exponents, we, indeed, observe anisotropy $(H_{yj}\approx 0.5)$
between longitudinal and transverse directions of multiparticle production, 
while there is an isotropy
in the transverse plane $(H_{ij}\approx 1$ for $i$ and $j$ both in the 
transverse plane). This result means that fractality in multiparticle 
production is self-affine rather than self-similar.

In order to show the independence of this conclusion of the 
particular set of variables
being used, the same analysis has also been done 
with the set ($y, p_{\ty}, p_{\tz}$) instead of ($y, p_\rt, \vf$). 
The corresponding results for the Hurst exponents are:
$$H_{yp_{\ty}}={1.0121\over 2.14}=0.47 \pm 0.06 \ ;$$
$$H_{yp_{\tz}}={1.0041 \over 2.14} = 0.47 \pm 0.06 \ ;$$
$$H_{p_{\ty}p_{\tz}} ={1.0041\over 1.0121}= 0.99 \pm 0.01 \ .$$
They show that the rule $H_{yj}\approx 0.5$, $H_{ij}\approx 1$ for 
$i$ and $j$ denoting variables in the transverse plane also holds for the 
variable set ($y, p_{\ty}, p_{\tz}$).

\section{Self-affine analysis}

With the Hurst exponents obtained above, we can perform a self-affine 
analysis in three- and two-dimensional phase space. For convenience, we
approximate the Hurst exponent for $(y,p_\rt)$ and $(y,\vf)$ 
by
$$\qquad\qquad \qquad H_{yj}=\frac{1}{2},\qquad \qquad (j=p_\rt, \vf),$$
but use
$$H_{p_\rt\vf}=1.\qquad \qquad $$
\noindent
From (3), it follows that 
$\la_{p_\rt}=\la_\vf=\la_y^2$. 
For $p_\rt$ and $\vf$, we use a 
partitioning
$M_y=1,2,3,\dots,10$ and 
$M_{p_\rt}=M_\vf=1,4,9,\dots,100$ in two-dimensional analysis, but 
$M_y=1,2,3,\dots,7$ and
$M_{p_\rt}=M_\vf=1,4,9,\dots,49$ in three-dimensional analysis. 
The results of the three-dimensional self-affine analysis
on $F_2$ are given by solid circles in
Fig.~3. Those of the corresponding self-similar analysis are 
repeated by open circles, for comparison. {In order to show the quality 
of the scaling law, linear fits 
\beq
\ln F_q=A+\phi_q \ln M_y 
\eeq
are compared to the data in Fig.~3. The fit results are given in Table 2.
To reduce the influence
of momentum conservation, the first point is not used in the fits.

Contributions to $\chi^2$ as shown in the top part of Table 2 are  
only from the diagonal terms of the covariance matrix. 
In fact, the diagonal terms provide the main contribution, 
and the relative size of $\chi^2$ for self-similar and self-affine
analyses should not change dramatically by adding the contribution 
from non-diagonal terms that account for the correlation between points at 
different bin size. As  is shown by comparing the results from fits to
the unweighted sample, with and without considering the
non-diagonal terms (two lower parts of Table 2), the fit results are
retained better
for the self-affine than for the self-similar analysis. A similar 
conclusion can be drawn from an inspection of Fig.~3, itself. While the
self-similar analysis leads to an upward bending, this effect is absent in 
the self-affine analysis. Of course, higher statistics data would be 
needed to definitively prove this point.

Even though the errors in the self-affine analysis are large, 
the results shown in Fig.~3
support the expectation [7] that a 3-dimensional self-affine analysis 
would lead to the full increase of $\ln F_q$ with increasing $\ln M$
right from the beginning. On the other hand, in a self-similar analysis,
the full increase would only be reached for large $M$, so that an upward
bending would be observed. This upward bending is indeed present for 
the self-similar analysis. If future experiments can confirm
the linear increase of the self-affine results with improved statistics,
this would mean that the scaling law (1) is better observed in 
self-affine analysis than in self-similar analysis.



The two-dimensional self-affine projections are presented in Fig.~4.
It can be seen that $\ln F_2$ 
increases smoothly with increasing $\ln M_y$
and the trends are similar for $(y,p_\rt)$ and $(y,\vf)$ (neglecting 
the first point in $(y,\vf)$),
meaning that in a self-affine analysis, the influence of artificial 
projection effects is reduced with respect to that observed in self-similar
analysis. }

\section{Conclusions}

In this paper we present a self-affine analysis of factorial moments in 
three-,
as well as in two-dimensional phase space, on the NA22 data for $\pi^+\Pp$ and 
$\PK^+\Pp$ collisions at 250 GeV/$c$. The results are compared with those
from a corresponding self-similar analysis.

From fitting the factorial moments of the one-dimensional projections by
the Ochs saturation formula (5), the Hurst exponents $H_{ij}$ are derived 
for all 
combinations of phase-space variables used. The value of $H_{ij}$ for a 
combination 
of rapidity $y$ with a transverse direction is approximately equal to 0.5.
We conclude, therefore, that fractality in multiparticle production is self-affine, rather than self-similar.
In the transverse plane, $H_{ij}$
stays approximately equal to 1.0 and, therefore, shows
merely self-similar fluctuation within that particular plane. Such a 
behavior can be understood from the privileged role of the longitudinal 
axis in multihadron production
and the symmetry within the plane transverse to this direction. 
This important point has been neglected in fluctuation analysis 
of multiparticle final states, so far.


Furthermore, the three-dimensional self-affine analysis shows a better scaling
behavior than does the corresponding self-similar analysis. The
two-dimensional 
self-affine projections with $H_{ij}=0.5$, i.e., $(y,p_\rt)$ and $(y,\vf)$
turn out to show a behavior more similar to each other than in the 
corresponding self-similar projections.


It would be interesting to see how the value of $H_{ij}$ changes with the
type of collision and with incident energy.

\vs 1cm

\ej
\ni
{\bf Acknowledgments}

We are grateful to the III. Physikalisches Institut B, RWTH
Aachen, Germany, the DESY-Institut f\"ur Hochenergiephysik, Berlin-Zeuthen,
Germany, the Department of High Energy Physics, Helsinki University, Finland,
the Institute for High Energy Physics, Protvino, Russia,
and the University of Warsaw and Institute of Nuclear Problems, Warsaw,
Poland, for early contributions to this experiment.Work is supported in part
by Polish KBN grant no. 2 P03B 083 
08 and by Polish-German Collaboration Foundation FWPN no. 1441/LN/94.
This work is part of the research program of the ``Stichting
voor Fundamenteel Onderzoek der Materie (FOM)", which is financially
supported by the ``Nederlandse Organisatie voor Wetenschappelijk Onderzoek
(NWO)". We further thank NWO for support of this project within the program
for subsistence to the former Soviet Union (07-13-038).
The work is also a part of the research project ``Density Fluctuations
in Multiparticle Production'', supported by the National Commission of
Science and Technology of China and the Koninklijke
Nederlandse Akademie van Wetenschappen (KNAW).
It is, furthermore, supported in part by the NNSF of China, the DYTF of the
State Education Commission of China and the CGP for young scientists
of Wuhan City.

\ej

\ni
{\bf REFERENCES}
\vs 5mm

\begin{itemize}
\item[1.] A. Bia\l as and R. Peschanski: Nucl. Phys. B273 (1986) 703 and B308 
(1988) 857.

\item[2.] See for example the review article: E. A. De Wolf,
I. M. Dremin and W. Kittel: Scaling Laws for Density Correlations and
Fluctuations in Multiparticle Dynamics, Nijmegen Preprint HEN-362 (1995),
Phys. Report (1996), in press.
\item[3.] B. Mandelbrot: The Fractal Geometry of Nature (Freeman, NY, 1982).
\item[4.] P. Lipa and B. Buschbeck: Phys. Lett. B223 (1989) 465.
\item[5.] W. Ochs: Phys. Lett. B247 (1990) 101.
\item[6.] N. Agababyan et al., NA22 Coll.: Z. Phys. C59 (1993) 405 and
Phys. Lett. B332 (1994) 458.
\item[7.] Wu Yuanfang and Liu Lianshou: Phys. Rev. Lett. 21 (1993) 3197. 
\item[8.] M. Aguilar-Benitez et al.: Nucl. Instrum. Methods 205 (1983) 79.
\item[9.] M. Adamus et al., NA22 Coll.: Z. Phys. C32 (1986) 475.
\item[10.] F. P. M. Botterweck: Ph.D. Thesis, University of Nijmegen, 1992.
\item[11.] B. Mandelbrot in Dynamics of Fractal Surfaces, eds. E. Family
and T. Vicsek (World Scientific, Singapore, 1991).
\item[12.] J. Wosiek: Proc. XXIV International Symposium on Multiparticle 
          Dynamics, Vietri sul Mare (Italy) 1994, eds. 
          A. Giovannini, S. Lupia and R. Ugoccioni (World 
          Scientific, Singapore, 1995) p. 99.
\item[13.] W. Ochs: Z. Phys. C50 (1991) 339. 
\item[14.] Liu Lianshou, Zhang Yang and Deng Yue: On the influence of 
momentum conservation upon the scaling behaviour of factorial moments 
in high energy multiparticle production, Wuhan preprint HZPP-9605
(1996), Z. Phys. C, in press.
\\
\end{itemize}

\ej

\vspace{5mm}
{\bf Table 1.} The parameter values obtained from a fit by (5).

$$\begin{array}{|c|c|c|c|c|}\hline
	  &  &    &                  &             \\
{\mbox{variables}}  & a  &  b & c & {\mbox{Omitting}} \\
                    &    &    &   & {\mbox{point(s)}} \\
\hline
   &            &         &                     &            \\

$$ y $$ & 1.336\pm  0.005 & 0.218  \pm 0.042 & 1.140 \pm 0.245
&  1 \\
$$ p_\rt$$   & 1.534\pm 0.021 & 0.340 \pm 0.021 & 0.021\pm 0.006
&  1 \\
$$ \vf $$   & 1.497\pm 0.019 & 0.420 \pm 0.019 & 0.014 \pm 0.005
&  1-3\\ 
\hline
\end{array}$$

\vspace{5mm}

{\bf Table 2.} The parameter values obtained from a fit by (7).

$$\begin{array}{|c|c|c|c|}\hline
 & \multicolumn{3}{c|}{{\mbox {without \ bin-size \ correlation}}}\\
\cline{2-4} 

   {\mbox{method}} & A  &  \phi_2 & \chi^2/NDF \\
\hline
   &                &               & \\
 {\mbox{weighted\ self-affine}} &  -0.04\pm 0.03 &  0.32\pm 0.03 & 7/4\\
 {\mbox{weighted\ self-similar}}&  ~0.11\pm 0.01  &  0.12\pm 0.01& 9/4\\
\hline
\hline
 & \multicolumn{3}{c|}{{\mbox {without \ bin-size \ correlation}}}\\
\cline{2-4} 
{\mbox{method}}  & A  &  \phi_2 & \chi^2/NDF \\
\hline
 &                   &             &\\
{\mbox{unweighted\ self-affine}} &  -0.08\pm 0.02 &  0.33\pm 0.03 & 12/4\\
 {\mbox{unweighted\ self-similar}}&  ~0.09\pm 0.01  &  0.10\pm 0.01& 20/4\\
\hline
\hline
 & \multicolumn{3}{c|}{{\mbox {with \ bin-size \ correlation}}} \\
\cline{2-4}
 {\mbox{method}}& A  &  \phi_2 & \chi^2/NDF \\
\hline
 &                   &             &\\
{\mbox{unweighted\ self-affine}} &  -0.08\pm 0.02 &  0.34\pm 0.02 & 14/4\\
{\mbox{unweighted\ self-similar}} &  ~0.10\pm 0.01 &  0.10\pm 0.01 & 32/4\\
\hline
\end{array}$$

\vspace{5mm}

\ej
\ni
{\bf FIGURE CAPTIONS}

\vspace{5mm}

\begin{itemize}
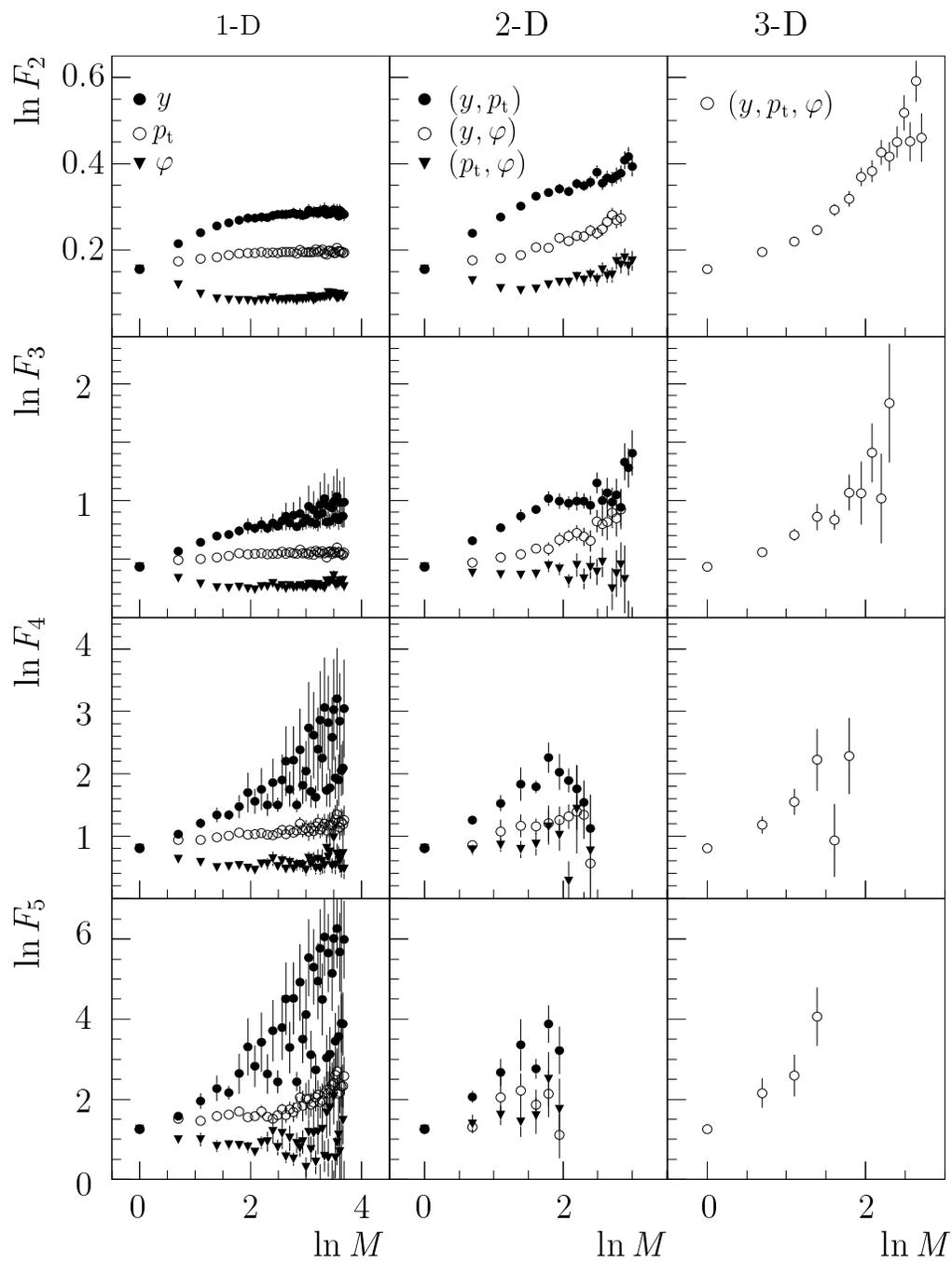
\item[Fig.~1] Self-similar analysis of $F_q$ in the set of variables
$(y,p_\rt,\vf)$ in one, two, and three dimensions, as indicated.
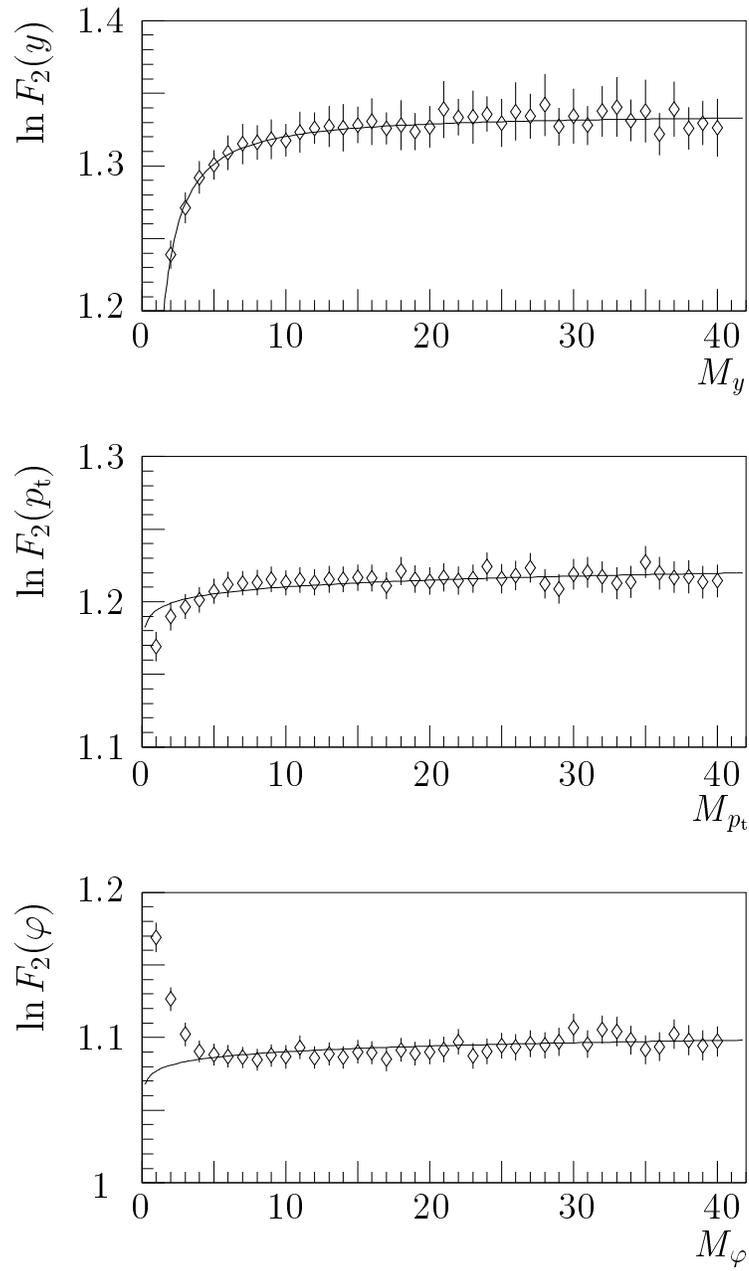
\item[Fig.~2]  Saturation curves for $F_2$ in the three one-dimensional
variables indicated. The curves are fits by (5) after omission of the
first point (first 3 points in the case of $F_2(\vf)$).
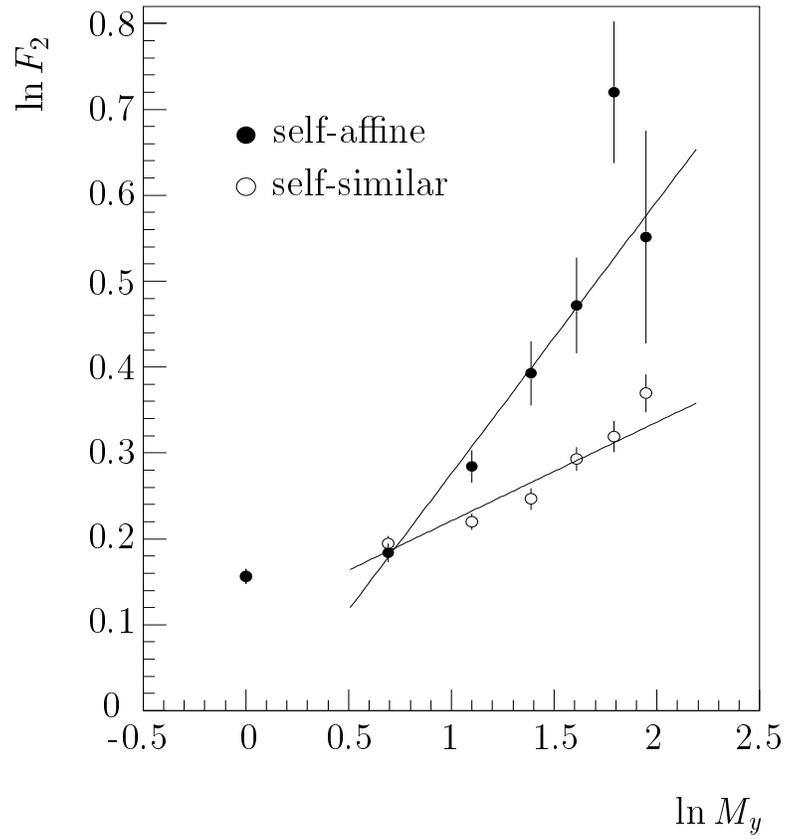
\item[Fig.~3]  Comparison of the three-dimensional self-affine
and self-similar analyses.
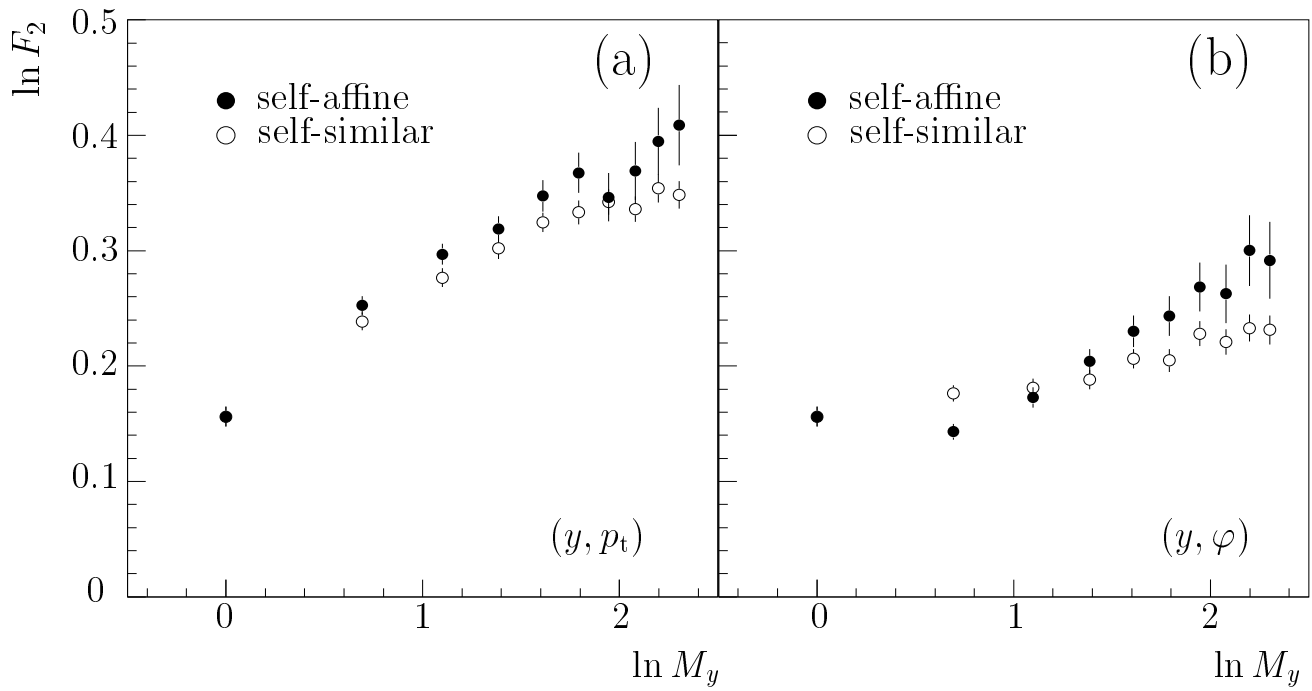
\item[Fig.~4]  Comparison of the two-dimensional projections of self-affine 
and self-similar analyses.
\end{itemize}

\end{document}